\documentclass{article}

\usepackage{iclr2027_conference,times}
\setcitestyle{numbers,square,comma}
\usepackage{fix-cm}
\usepackage[T1]{fontenc}    %
\usepackage{hyperref}       %
\usepackage{url}            %
\usepackage{booktabs}       %
\usepackage{amsfonts}       %
\usepackage{nicefrac}       %
\usepackage{microtype}      %
\usepackage{xcolor}         %
\usepackage{listings}
\usepackage{xspace}
\usepackage{enumitem}
\usepackage{amsmath}
\usepackage{cleveref}
\usepackage{graphics}
\usepackage{graphicx}
\usepackage{multirow}

\makeatletter
\AddToHook{cmd/appendix/before}{\def\cref@section@alias{appendix}}
\makeatother

\usepackage[most]{tcolorbox}
\newtcolorbox{defbox}[1]{
  enhanced,
  breakable,
  colback=gray!3,
  colframe=black!55,
  boxrule=0.4pt,
  arc=2pt,
  left=10pt, right=10pt, top=6pt, bottom=6pt,
  fonttitle=\bfseries,
  coltitle=black,
  title=#1,
}

\definecolor{keywordcolor}{rgb}{0.7, 0.1, 0.1}   %
\definecolor{tacticcolor}{rgb}{0.0, 0.1, 0.6}    %
\definecolor{commentcolor}{rgb}{0.4, 0.4, 0.4}   %
\definecolor{symbolcolor}{rgb}{0.0, 0.1, 0.6}    %
\definecolor{sortcolor}{rgb}{0.1, 0.5, 0.1}      %
\definecolor{attributecolor}{rgb}{0.7, 0.1, 0.1} %

\newcommand{\code}[1]{\lstinline|#1|}

\newcommand{\aqueduct}{\textsc{Aqueduct}\xspace}
\newcommand{\verina}{Verina\xspace}

\title{Agentic Verified Optimization of Real-World Software}

\author{%
Zhe Ye\textsuperscript{1}\thanks{Work done during internship at Amazon Web Services.} , Udaya Ghai\textsuperscript{2}, Zhizhen Qin\textsuperscript{2}, Dawn Song\textsuperscript{1}, Soonho Kong\textsuperscript{2} \\
\textsuperscript{1}University of California, Berkeley, \textsuperscript{2}Amazon Web Services
}

\iclrfinalcopy

\AddToHook{cmd/maketitle/after}{\fancyhead{}}

\begin{document}

\maketitle

\begin{abstract}
Program optimization aims to improve performance while preserving behavior.
Coding agents can already make programs faster by changing algorithms, data structures, and execution strategies.
However, existing equivalence checks mostly rely on tests, which agents have been shown to exploit, passing the tests without preserving the intended behavior.
We propose \emph{agentic verified optimization}, in which an agent produces a faster implementation together with a machine-checked proof of equivalence to the existing one.
Because the existing implementation serves as the specification, the task avoids the specification-writing bottleneck of verified code generation, and recent advances in theorem proving make the proofs attainable.
We formulate the task at the function and repository levels in Lean~4, where programs can be both executed and reasoned about.
Few real-world applications are written in Lean, so we develop \aqueduct{}, an agentic framework that migrates C, C++, Go, and Rust programs into native Lean, validates each port by differential testing against its original.
On the 189 reference implementations in \verina{}, agents produce faster proved rewrites for 156, including algorithmic changes such as replacing exponential recursion with dynamic programming.
On 17 migrated applications, their 92 proved rewrites speed up 47 of 51 application scenarios.
These application-level gains are smaller than the function-level ones, making the choice of functions to optimize an important problem.
Current agents can thus already carry out verified optimization on both standalone functions and applications migrated from real programs.
We discuss the remaining challenges and the steps toward verified optimization in production.
\end{abstract}

\section{Introduction}
\label{sec:intro}

Coding agents are increasingly used to write and modify software~\citep{chen2021evaluating,jimenez2024swe,yang2024sweagent}.
They are now also asked to make code faster~\citep{huang2024effibench,du2024mercury,waghjale2024ecco}.
This goal continues a long line of work on automatic optimization.
Before language models, superoptimizers searched instruction sequences~\citep{massalin1987superoptimizer,schkufza2013stochastic}, learned models guided search over programs~\citep{bunel2017learning}, and reinforcement learning discovered faster algorithms~\citep{mankowitz2023faster}.
Language models and coding agents have since taken on the task, as PIE~\citep{shypula2024learning} rewrites programs for speed, KernelBench~\citep{ouyang2025kernelbench} studies efficient GPU kernel generation, and Jitskit~\citep{liu2026time} synthesizes systems specialized to a workload.
These optimizations range from local changes that eliminate redundant computation to new data representations and algorithms.
Whatever the strategy, a faster replacement must preserve the behavior of the program it replaces.

Agents can already find substantial performance gains, but existing work checks behavior preservation with tests, which, as Dijkstra~\citep{dijkstra1972humble} observed, can show the presence of errors but not their absence.
This limitation matters more when the tests also define an agent's reward.
Agents then learn to pass the tests rather than preserve the behavior, a pattern studied as reward hacking~\citep{skalse2022defining} and observed in coding agents that special-case the tested inputs~\citep{zhong2026impossiblebench}.
In optimization, reward hacking yields speedups from programs that do not preserve behavior.
Jitskit~\citep{liu2026time} describes an apparent $6\times$ improvement that came from regenerating test values rather than storing them, and an audit found correctness failures in 39.5\% of the GPU kernels that an optimization harness had accepted~\citep{shah2026contract}.
A proof of equivalence avoids this problem because it covers every input, the criterion used in translation validation~\citep{pnueli1998translation,necula2000translation} and regression verification~\citep{godlin2013regression,churchill2019semantic}.
With a proof as the check, an agent can change algorithms and data representations freely without weakening correctness.

Agents are now capable provers.
Neural theorem proving has progressed from language-model proof search~\citep{polu2020generative,lample2022hypertree} to provers trained with reinforcement learning that reach olympiad level~\citep{hubert2026olympiad,ren2025deepseek,lin2026goedel}, and agents now formalize research mathematics, from open Erd\H{o}s problems~\citep{sothanaphan2026resolution,tsoukalas2026advancing} to a complete Lean proof of Fermat's Last Theorem~\citep{chen2026prove2me,anthropic2026formalizing}.
They also build verified software in Lean, from repository-level implementations with proofs~\citep{ye2026vero} to a verified compiler~\citep{rinard2026testing} and a verified DEFLATE library~\citep{morrison2026leanzip}.
Program optimization is a natural application of these capabilities.

We study \emph{agentic verified optimization}, in which an agent produces a faster implementation $f'$ of a reference function $f$ together with a machine-checked equivalence proof.
The proof establishes $\forall x,\, f(x)=f'(x)$, with the reference implementation $f$ as the specification for correctness.
This avoids the specification-writing problem in verified code generation, where writing specifications that capture the intended behavior remains difficult, even manually~\citep{endres2024can,ye2026intent,ma2025specgen}.
This task can also be viewed as the refinement step that formal methods have long advocated, in which a simple implementation is proved correct and then refined into a fast one that inherits its guarantees.
By hand, it is costly and has been reserved for critical domains such as cryptography~\citep{erbsen2019simple}.
Agents are a natural fit for this work.
In verified code generation, proofs can push agents toward simple implementations~\citep{ye2026vero}.
Agentic verified optimization keeps the simple, proved version and then allow agent to search complicated, optimized version with the same correctness guarantee.
The task in turn helps improve agents.
Every function in an existing codebase is an instance with its specification in place, and because correctness is fixed while speedup is a continuous measure rather than a pass rate, the task is open-ended and does not saturate as agents improve.

The task couples implementation and proof search.
For example, replacing recursive enumeration with dynamic programming can remove repeated work, but its proof needs an invariant connecting table entries to the original recursion.
An implementation may be equivalent and fast yet difficult for the agent to prove within a limited search budget.
Conversely, a readily proved rewrite may offer little improvement.
The agent must decide whether to improve a candidate further, invest in its proof, or pursue another implementation.
Verified optimization therefore tests how agents coordinate program reasoning, performance analysis, and proof construction.

We formulate the task in Lean~4~\citep{moura2021lean4}, where the same definitions can be executed and used in proofs, and whose compiler produces native code with a reference-counting runtime~\citep{ullrich2019counting}.
Lean's kernel checks each equality proof, and execution on the evaluation workloads measures the speedup.
We study current agent capabilities at two levels.
\emph{Function-level optimization} asks the agent to improve a given implementation.
We use the 189 reference implementations in \verina{}~\citep{ye2026verina} to study this setting.
\emph{Repository-level optimization} asks the agent to improve an application by choosing and rewriting functions within its codebase.
Here, function-level microbenchmarks measure individual rewrites, while application-level macrobenchmarks measure whether they speed up the application.
This setting requires executable applications in Lean, but few real-world applications are written in Lean.
We therefore build \aqueduct{}, an agentic framework that migrates C, C++, Go, and Rust programs into native Lean and checks each port against its original through differential testing~\citep{mckeeman1998differential}.
The resulting ports keep the interfaces, behavior, and workloads of real software and serve as targets for verified optimization.

On \verina{}, agents produce 156 verified improvements among the 189 targets.
Using the fastest verified implementation for each target, the geometric mean speedup is $1.851\times$ and the median is $1.117\times$.
On 17 migrated repositories, agents produce 92 rewrites proved equivalent to the Lean ports they replace, all faster on function-level microbenchmarks.
The geometric mean speedup is $1.864\times$, with a median of $1.402\times$.
Fifteen of these optimized functions outperform functions in the original source-language programs.
On 51 application scenarios, execution improves in 47 cases with a geometric mean speedup of $1.120\times$ over the migrated Lean ports.
The repository rewrites remove repeated normalization and temporary lists, add fast paths for common input cases, and change traversal order while preserving the reference behavior.
Together, these findings outline concrete steps toward verified optimization of real-world software, which we discuss in Section~\ref{sec:future}.

\section{Background and Related Work}
\label{sec:related}

Existing work on program optimization and verification shares several limitations that prevent agents from optimizing real-world software with guaranteed correctness.

\textbf{First, checking optimization correctness with tests does not guarantee behavior preservation.}
PIE~\citep{shypula2024learning} generates program edits for speed, and LLM Compiler~\citep{cummins2025llm} selects compiler optimizations for code size.
At the algorithm level, AlphaDev~\citep{mankowitz2023faster} and AlphaEvolve~\citep{novikov2025alphaevolve} search for improved algorithms.
For GPU kernels, generation and optimization are studied in KernelBench~\citep{ouyang2025kernelbench}, Sakana's CUDA Engineer~\citep{lange2025towards}, CUDA-L1~\citep{li2026cuda}, Astra~\citep{wei2025astra}, and AsmEvo~\citep{liu2026asmevo}.
At system scale, Jitskit~\citep{liu2026time} synthesizes systems from specifications.
These systems check correctness with tests or differential execution.
When this assessment determines the reward for optimization, an agent can improve its score by exploiting the checks or omitting work needed outside the tested cases.
Documented failures include exploiting harness bugs or weaknesses in timing methodology~\citep{lange2025towards,li2026cuda}, specializing to benchmark input distributions~\citep{liu2026time,zhang2026kernelbench}, and passing standard tests while violating kernel contracts~\citep{sarkar2026correctness,shah2026contract}.
In agentic verified optimization, an equivalence proof rules out these correctness failures because it covers every input.

\textbf{Second, optimizers built on predefined transformations or representations cannot go beyond them.}
Verified compilers such as CompCert~\citep{leroy2009formalverification} and CakeML~\citep{kumar2014cakeml} apply passes proved correct in advance, and egg~\citep{willsey2021egg} explores the consequences of supplied rewrite rules.
Superoptimizers~\citep{massalin1987superoptimizer} such as Denali~\citep{joshi2002denali}, STOKE~\citep{schkufza2013stochastic}, and Souper~\citep{sasnauskas2018souper} search short instruction sequences within a fixed instruction representation.
At the source level, Fiat~\citep{delaware2015fiat}, the Isabelle Refinement Framework~\citep{lammich2012applying}, and Cogent~\citep{oconnor2021cogent} derive implementations from refinement steps supplied by a developer or compiler, and LLMLift~\citep{bhatia2024verifiedcode} verifies translations of code into domain-specific languages such as Spark and TACO, so any performance benefit comes from the optimizations those languages already provide.
Even when every transformation is verified, such an optimizer finds only what its passes, rules, representation, or target language can express, which places a ceiling on its gains.
An agent has no such fixed set, since it can write a new algorithm or data representation directly in Lean and prove it equivalent to the reference without moving the program to another language.

\textbf{Third, agents that optimize under formal checks have limited ability to prove their rewrites.}
Translation validation~\citep{pnueli1998translation,necula2000translation} checks each compiler output with an automated equivalence checker, as Alive2~\citep{lopes2021alive2} does for LLVM with bounded loop unrolling and HEC~\citep{yin2025hec} does with equality saturation.
LLM-Vectorizer~\citep{taneja2025llm}, VecTrans~\citep{zheng2025vectrans}, and LLM-VeriOpt~\citep{fang2026llm} check their LLVM rewrites with Alive2, which accepts only 38.2\% of LLM-Vectorizer's test programs, and T-LLM~\citep{fazel2026tllm}, which pairs the bounded model checker CBMC with a language-model judge, reports a 33.8\% false-acceptance rate.
AutoGate~\citep{wang2026autogate} and Ares~\citep{cuyckens2026ares} check hardware rewrites with commercial equivalence tools, CertiBlock~\citep{si2026certificate} checks six predefined rewrite families for Scratch programs, and CktFormalizer~\citep{xiong2026cktformalizer} proves equivalence for circuits in a domain-specific language embedded in Lean.
Other systems let the agent prove its rewrite against a written specification, such as IDS~\citep{agarwal2026inductive} for distributed systems, Solidus~\citep{paradigm2026soliduscompiler} for a verified Yul-to-EVM compiler, and lean-zip~\citep{morrison2026leanzip} for a verified DEFLATE implementation.
These specifications are hard to write and must be written anew for each system~\citep{endres2024can,ye2026intent}.
Our task uses the existing implementation as the specification, so no specification has to be written before a function can be optimized.

\textbf{Fourth, verified code-generation benchmarks measure correctness but not performance.}
Function-level benchmarks in Lean, including miniCodeProps~\citep{lohn2024minicodeprops}, FVAPPS~\citep{dougherty2025provingcoding}, \verina{}~\citep{ye2026verina}, and CLEVER~\citep{thakur2025cleververified}, benchmarks for SMT-based verifiers such as DafnyBench~\citep{loughridge2025dafnybench} and VerusBench~\citep{chen2025automated}, VeriCoding~\citep{chiosa2026benchmark} across Dafny, Verus, and Lean, and repository-scale benchmarks such as RVBench~\citep{zhong2025towards}, VeruSAGE-Bench~\citep{yang2026verusage}, VeriSoftBench~\citep{xin2026verisoftbenchcolm}, and Vero~\citep{ye2026vero} all check whether an implementation satisfies a specification, and none evaluates its speed.

\textbf{Finally, Lean has few executable real-world applications to optimize.}
Lean is both a proof assistant and a general-purpose language with a native compiler~\citep{moura2021lean4,ullrich2019counting}, but its software developments are formal models of languages, such as Aeneas~\citep{ho2022aeneas} and Strata~\citep{strata2026strata}, and individual components, such as Cedar~\citep{cutler2024cedarauthorization}, lean-zip~\citep{morrison2026leanzip}, and RyuLean4~\citep{ryulean2026ryulean4verification}, none of which has the callers and workloads of a deployed application.
Language models can translate programs between languages~\citep{jiang2026large}, with translations checked by bounded model checking in VERT~\citep{yang2025vert} or by differential fuzzing in FLOURINE~\citep{eniser2025towards}.
\aqueduct{} migrates applications into native Lean and validates each port with differential testing~\citep{mckeeman1998differential}, coverage-guided fuzzing~\citep{zalewski2014americanfuzzy,llvm2026libfuzzerlibrary}, and mutation analysis~\citep{jia2011analysis}.

\section{Agentic Verified Optimization}
\label{sec:task}

\begin{figure}[t]
\centering
\includegraphics[width=\linewidth]{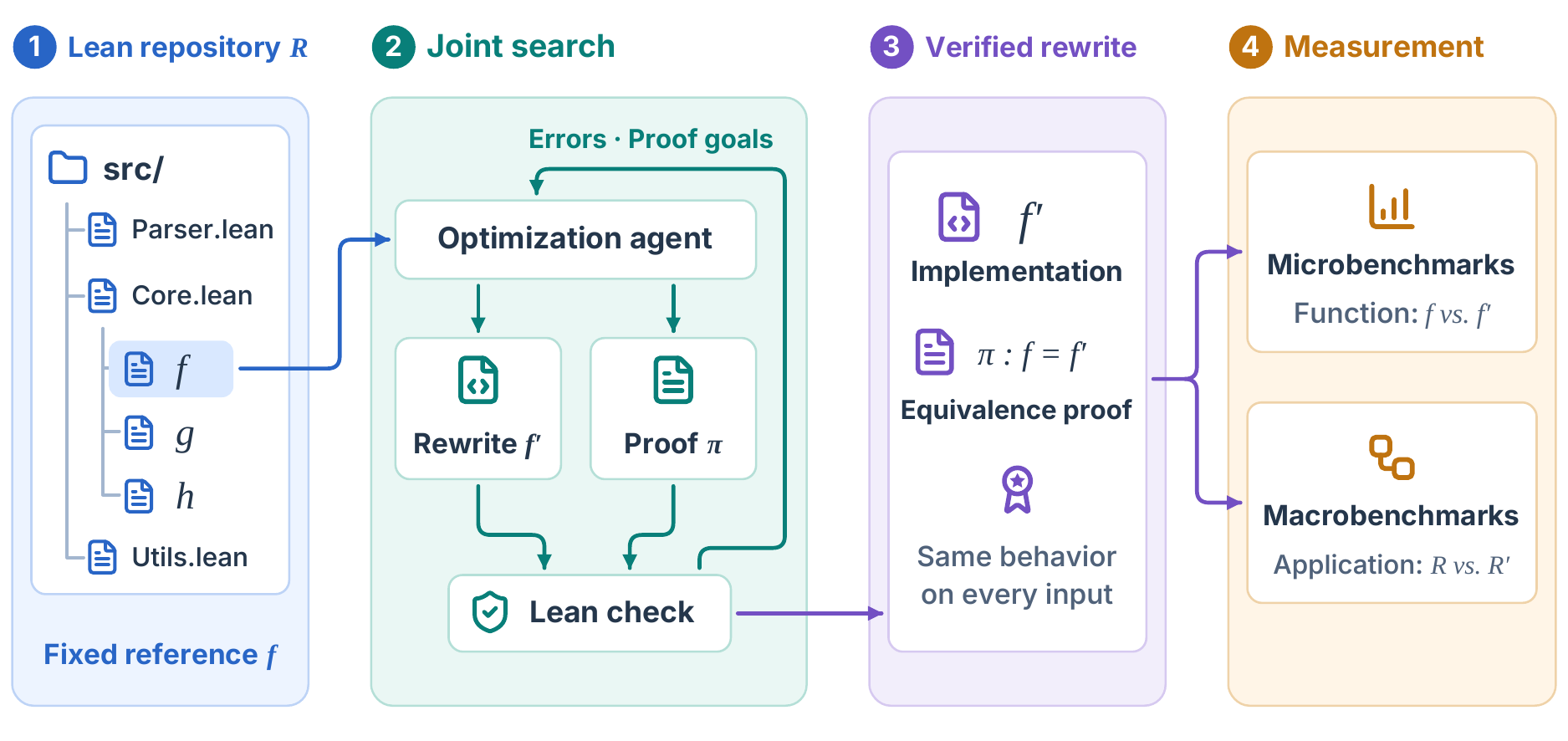}
\caption{\textbf{Agentic verified optimization in a repository.} An agent selects a function $f$ from a Lean repository and jointly searches for a replacement $f'$ and an equivalence proof $\pi$. Keeping $f$ fixed, it uses Lean's errors and proof goals to revise both artifacts. The search returns the replacement with a checked proof that it preserves the reference's behavior. Microbenchmarks measure function-level speedup, and macrobenchmarks measure the effect on the application.}
\label{fig:workflow}
\end{figure}

The task starts with an executable reference and asks an agent to improve its performance while preserving its behavior.
For a given function, this requires searching for an implementation together with its equivalence proof.
Within a repository, the agent must also choose which functions to optimize (Figure~\ref{fig:workflow}).

\subsection{Problem Formulation}
\label{sec:formulation}

Program optimization can be expressed as a search for a faster implementation under an equivalence constraint.
STOKE~\citep{schkufza2013stochastic} uses this formulation to search instruction sequences, approximating equivalence with tests during search and validating the result afterward.
We apply the same formulation to source-level programs, with the agent searching for both a replacement and its equivalence proof.

\paragraph{Behavioral equivalence.}
A reference implementation fixes the behavior that an optimization must preserve.
For a pure function, this means returning the same result on every input.
For a function whose effects are expressed inside Lean, such as state threaded through a monad or exceptions, the effects are part of the function's value and the same equality applies.
In-place updates of uniquely referenced arrays are ordinary pure code in Lean, so representation changes such as replacing lists with arrays or preallocating buffers fall within the task.
Opaque effects, such as input and output or foreign calls, whose behavior Lean does not define, fall outside this study.
Equality can also be stronger than callers need, for example when using a floating-point result within a tolerance.
Section~6 discusses modelling opaque effects and replacing equality with a weaker relation.
We express this obligation in Lean~4 for an executable reference function $f$ in a repository $\mathcal{R}$.
The replacement $f'$ has the same signature.
Its equivalence certificate is a proof $\pi : f = f'$.
By function extensionality, the agent can establish it by proving $\forall x,\, f(x)=f'(x)$.
The proof permits replacing calls to $f$ with calls to $f'$ while preserving their results.

\paragraph{Equivalence instead of a written specification.}
Verified code generation usually states correctness against a formal specification written for the task.
Joint implementation-and-proof systems such as IDS~\citep{agarwal2026inductive} and Vero~\citep{ye2026vero} prove that an implementation refines or satisfies such a specification, and an optimizer could in principle be scored against one as well.
A specification, however, must first be written and then trusted.
Writing one that captures the intended behavior is difficult~\citep{endres2024can,ye2026intent,ma2025specgen}, and manually curated specifications contain errors~\citep{ye2026vero}.
A specification that omits part of the intended behavior also lets an agent satisfy it while dropping that behavior, the same reward hacking that weak tests invite~\citep{skalse2022defining}.
Optimization has a simpler starting point.
The reference implementation is executable, so its behavior can be read and tested directly, and it can be written as the simplest correct algorithm regardless of speed.
Equality to it specifies the behavior completely, and the obligation does not change as the agent explores faster algorithms.
This is also the setting of regression verification, in which the previous version of a program serves as its specification~\citep{godlin2013regression}.

\paragraph{Optimization objective.}
The agent receives the reference and its repository context and searches for a replacement together with a proof.
The evaluator fixes a workload $\mathcal{B}$ on which speed is measured.
Writing $t_{\mathcal{B}}$ for execution time on that workload, the joint objective is
\begin{equation}
\underset{f',\,\pi}{\operatorname{minimize}}\quad t_{\mathcal{B}}(f')
\qquad\text{subject to}\qquad \pi : f = f'.
\label{eq:task}
\end{equation}
Here the constraint requires Lean's kernel to accept the proof with the reference fixed.
The replacement must be executable and build with its callers, and the agent may supply auxiliary definitions and lemmas needed by either artifact.
Section~\ref{sec:acceptance} specifies these checks.
Within a finite search budget, a successful result is a feasible pair with $t_{\mathcal{B}}(f') < t_{\mathcal{B}}(f)$.

\paragraph{Function-level optimization.}
The agent is given a target $f$ and seeks a faster implementation with an accepted equivalence proof.

\paragraph{Repository-level optimization.}
The agent receives the repository and chooses which functions to optimize within its search budget (Figure~\ref{fig:workflow}).
Each replacement requires an equivalence proof, and the integrated replacements must build with the application.
The objective is to reduce application execution time.
We write $\mathcal{R}'$ for the repository after integrating the accepted replacements.
Function-level microbenchmarks measure individual rewrites, while application-level macrobenchmarks compare $\mathcal{R}'$ with $\mathcal{R}$ (Section~\ref{sec:measurement}).
These measurements distinguish a locally faster function from an improvement to the application that uses it.

\paragraph{Target selection.}
Target selection determines how repository-level optimization spends its search budget.
A function is worth optimizing when it accounts for a large share of application time and its rewrite appears provable.
Source code, call structure, and profiles on development workloads indicate the first, and the function's structure and the lemmas already available indicate the second.
Targets also differ in granularity.
A local rewrite replaces one function, carries one equality obligation, and can gain at most that function's share of runtime.
A rewrite of a module or of a shared data representation changes several functions at once, and the obligation then falls on the module's interface, where each exported function must remain equal to its reference while internal helpers may change freely.
This holds when the new representation stays inside the module, and exposing it in exported types requires a refinement relation between the old and new types instead of equality.
The larger obligation gives access to representations and control paths that a rewrite of a single function cannot reach.
An agent can therefore spend its budget on many cheap local rewrites, on one structural rewrite with a long proof, or on a sequence in which local rewrites establish lemmas that a later structural rewrite reuses, and Section~\ref{sec:future} lists the comparisons that would distinguish these strategies.

\paragraph{Workloads and specialization.}
The workload $\mathcal{B}$ defines what faster means, and a study can withhold it, provide a sample of it, or describe the kind of input the function serves.
With test-based acceptance, revealing inputs is unsafe.
Tests establish correctness only on the inputs they contain, so an agent that sees them can special-case those inputs, for example by storing their outputs, and pass while computing wrong results elsewhere~\citep{liu2026time,zhang2026kernelbench}.
An equality proof covers every input, so a revealed sample can affect which inputs run fast but not whether the replacement is correct.
Withholding $\mathcal{B}$ measures improvements that generalize across inputs, and describing it lets the agent specialize, as Jitskit~\citep{liu2026time} does when it synthesizes a system for a declared workload.
A specialized implementation $g$ needs to be correct only on that kind of input, so it can assume properties such as bounded sizes or sorted lists, and use optimizations that a replacement correct on every input cannot.
The agent states the kind of input as a decidable predicate $P$ and proves $\forall x,\, P(x) \rightarrow f(x) = g(x)$.
The guarded replacement $h$, defined by $h(x) = g(x)$ if $P(x)$ and $h(x) = f(x)$ otherwise, then satisfies the full obligation $h = f$, at the cost of evaluating $P$ on each call.
If the callers guarantee $P$, the agent can deploy $g$ alone, and the same lemma certifies it on every input satisfying $P$.
In both cases the correctness argument is machine-checked, so the degree of specialization is a parameter of the study rather than a limit of the task.
Section~\ref{sec:eval-setup} states the choice made in our study, and Section~\ref{sec:future} discusses specialization as a direction.

\subsection{Joint Implementation and Proof Search}
\label{sec:design}
\label{sec:pipeline}

Each rewrite changes the proof the agent must write.
Hoisting a repeated computation requires showing that its value can be reused, a changed traversal requires an invariant over the elements already processed, and a new algorithm requires relating different intermediate states.
The agent therefore searches for the implementation and its proof together (Figure~\ref{fig:workflow}).
After proposing a rewrite, it can optimize further, attempt the proof, prove a supporting lemma, or change the implementation to make the proof easier, guided by Lean's errors and proof goals.
A failed proof does not tell the agent whether the rewrite is wrong or the proof is incomplete, so dividing effort between implementation and proof is part of the task.

\begin{figure}[t]
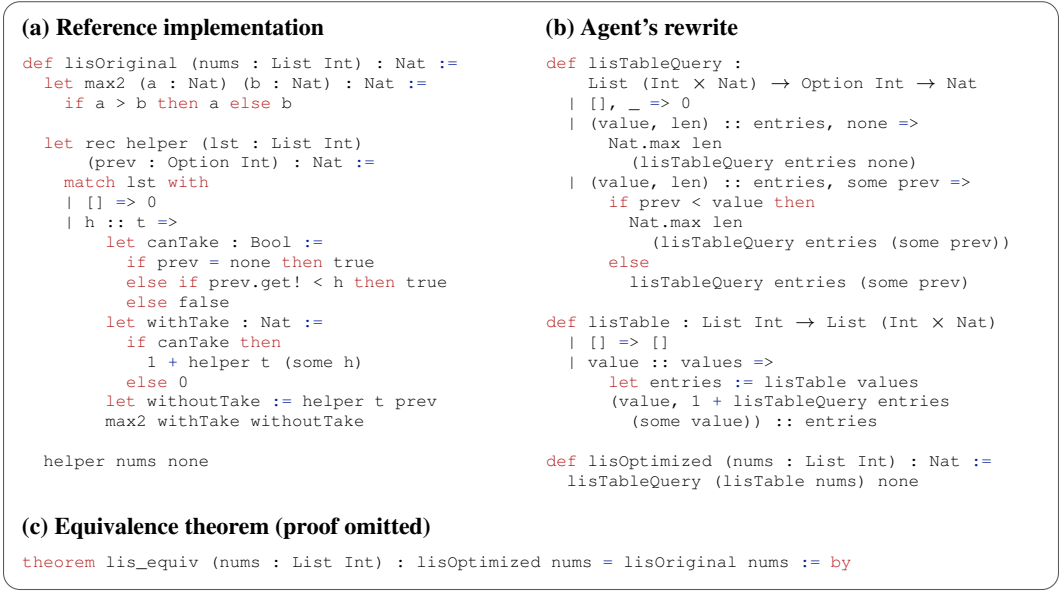

\lstset{language=lean,basicstyle=\ttfamily\fontsize{6.5}{7.5}\selectfont,numbers=none,xleftmargin=0pt,aboveskip=5pt,belowskip=0pt,breaklines=false}
\begin{tcolorbox}[colback=white,colframe=black!65,boxrule=0.4pt,arc=6pt,boxsep=0pt,left=7pt,right=7pt,top=6pt,bottom=6pt]
\noindent\begin{minipage}[t][184pt][t]{0.485\linewidth}
\vspace{0pt}
{\footnotesize\bfseries (a) Reference implementation}
\begin{lstlisting}[firstline=5,lastline=25]
def lisOriginal (nums : List Int) : Nat :=
  let max2 (a : Nat) (b : Nat) : Nat :=
    if a > b then a else b

  let rec helper (lst : List Int)
      (prev : Option Int) : Nat :=
    match lst with
    | [] => 0
    | h :: t =>
        let canTake : Bool :=
          if prev = none then true
          else if prev.get! < h then true
          else false
        let withTake : Nat :=
          if canTake then
            1 + helper t (some h)
          else 0
        let withoutTake := helper t prev
        max2 withTake withoutTake

  helper nums none
\end{lstlisting}
\end{minipage}\hfill
\begin{minipage}[t][184pt][t]{0.485\linewidth}
\vspace{0pt}
{\footnotesize\bfseries (b) Agent's rewrite}
\begin{lstlisting}[firstline=27,lastline=48]
def lisTableQuery :
    List (Int × Nat) → Option Int → Nat
  | [], _ => 0
  | (value, len) :: entries, none =>
      Nat.max len
        (lisTableQuery entries none)
  | (value, len) :: entries, some prev =>
      if prev < value then
        Nat.max len
          (lisTableQuery entries (some prev))
      else
        lisTableQuery entries (some prev)

def lisTable : List Int → List (Int × Nat)
  | [] => []
  | value :: values =>
      let entries := lisTable values
      (value, 1 + lisTableQuery entries
        (some value)) :: entries

def lisOptimized (nums : List Int) : Nat :=
  lisTableQuery (lisTable nums) none
\end{lstlisting}
\end{minipage}

\smallskip
\noindent{\footnotesize\bfseries (c) Equivalence theorem (proof omitted)}
\begin{lstlisting}[firstline=50,lastline=50]
theorem lis_equiv (nums : List Int) : lisOptimized nums = lisOriginal nums := by
\end{lstlisting}
\end{tcolorbox}
\caption{\textbf{From recursive search to dynamic programming.} A simplified example from \verina{} task \code{verina\_advanced\_33}, showing (a) the reference implementation, (b) the agent's rewrite, and (c) their equivalence theorem, proved in Appendix~\ref{app:lis-proof}. The rewrite runs $4{,}467.871\times$ faster on the 261-input corpus.}
\label{fig:task-format}
\end{figure}

Figure~\ref{fig:task-format} shows an example from \verina{}~\citep{ye2026verina}.
The reference for \code{longestIncreasingSubsequence} (\code{verina\_advanced\_33}) makes two recursive calls per element, one taking it and one skipping it, and so explores $2^n$ branches on an increasing list of length $n$.
The agent's rewrite fills a table from the end of the list with the length of the longest increasing subsequence starting at each position, visits $O(n^2)$ entries, and runs more than $4{,}000\times$ faster.
The proof is the hard part, because the reference recurses on the remaining list and the previously chosen element, while the table is indexed by position.
The agent bridged the two with a lemma relating each table query to the reference's recursive helper for every suffix and every possible preceding element, which makes the induction work for both the take and the skip branch.

Lemmas can also be reused across rewrites.
In Gron, a migrated JSON tool (Appendix~\ref{app:casestudies}), the proof of the string-validation rewrite \texttt{validIdentifier} reuses character-classification lemmas proved for an earlier rewrite.

\subsection{Acceptance and Measurement}
\label{sec:acceptance}
\label{sec:measurement}

A rewrite is accepted when Lean's kernel checks its equality theorem against the unchanged reference.
Following Vero~\citep{ye2026vero}, the check runs in a fresh copy of the repository, so the agent's declarations cannot alter the reference, add axioms, or substitute compiled code for the definition being proved (Appendix~\ref{app:acceptance}).
The proof establishes equality in Lean's logic, and as for any Lean program, the compiler and runtime are trusted to preserve it in the executable.

\paragraph{Metrics.}
A function-level microbenchmark times a target in isolation, and an application-level macrobenchmark, which we call an \emph{application scenario}, runs a public entry point on fixed inputs with all accepted rewrites integrated.
Function-level speedup is $s(f') = t_{\mathcal{B}}(f) / t_{\mathcal{B}}(f')$, application-level speedup for a scenario $w$ is $t_w(\mathcal{R}) / t_w(\mathcal{R}')$, and we summarize both with geometric means and medians.
For migrated applications, we also time the matching operation in the original program on the same inputs, which compares the ports with their native originals.
To describe how gains are distributed over a fixed set of targets $F$, let $s_f$ be the best verified speedup of target $f$, or $0$ if it has no verified candidate.
Following the $\mathrm{fast}_p$ metric of KernelBench~\citep{ouyang2025kernelbench}, we define
\begin{equation}
\mathrm{vfast}_p = \frac{1}{|F|} \sum_{f \in F} \mathbf{1}[s_f > p],\qquad p\geq 1,
\label{eq:vfast}
\end{equation}
where $|F|$ is the number of targets and the indicator $\mathbf{1}[s_f > p]$ equals $1$ when $s_f > p$ and $0$ otherwise.
Thus $\mathrm{vfast}_p$ is the fraction of targets whose best verified candidate is more than $p$ times faster than the reference, and $\mathrm{vfast}_1$ is the fraction with any verified improvement.

\section{Targets for Verified Optimization}
\label{sec:testbeds}

Both settings need executable Lean programs to optimize.
For function-level optimization, we use the standalone functions in \verina{}~\citep{ye2026verina} (Section~\ref{sec:verina-opt}).
Repository-level optimization additionally needs complete applications with callers and workloads, which Lean lacks (Section~\ref{sec:related}), so we build \aqueduct{}, an agentic framework that migrates C, C++, Go, and Rust programs into Lean and validates each port against its original (Section~\ref{sec:aqueduct}).

\subsection{Functions from \verina{}}
\label{sec:verina-opt}

\paragraph{Source and construction.}
\verina{} contains 189 standalone Lean~4 tasks whose reference implementations prioritize correctness and mostly use lists.
We use every reference implementation as $f$ and ask for a faster $f'$ with a proof $\pi : f=f'$ over the valid inputs that each task's precondition defines, without using \verina{}'s formal specifications, since the reference implementation is the specification.

\paragraph{Workloads.}
We use a deterministic generator to construct up to 256 cases per target and also include the official tests.
Because the functions are standalone, they are measured with microbenchmarks only.

\subsection{Applications Migrated by \aqueduct{}}
\label{sec:aqueduct}

\aqueduct{} obtains Lean applications by migrating existing programs, which gives them three properties that a freshly written corpus would lack.
Their code structure, configuration surface, and workloads come from deployed software rather than from a benchmark author.
The original program is an executable oracle, so a port can be validated on any input without a written specification, just as the reference implementation later specifies the optimization.
The original also serves as a performance baseline for the optimized Lean code.
The difficulty is fidelity.
A port that compiles and passes the common cases can still drop rare options, boundary inputs, exact diagnostics, and exit conventions, and tests written by the porting agent share its assumptions.
\aqueduct{} therefore develops the port and its test corpus together, and no agent judges its own work.

\paragraph{Migration task.}
Each task supplies a pinned original program and a declared input domain for its public interface.
A Lean port is compared with the original at its library API or executable entry points, which leaves the port's internal structure to the migration agents.
For each input, the comparison checks exit status, output bytes, filesystem effects, and timeout state, using byte equality by default and a declared tolerance only for output that cannot be stabilized, such as floating-point formatting.
The port must be native Lean with no foreign-function interface (FFI) or subprocess call to the original, since agreement is meaningless if the port consults the oracle.
Migration succeeds when the port agrees with the original on every test in the accumulated corpus $T$.
We report recovery, the fraction of tests passed by the original that the port also passes, measured once on a disjoint hidden test set $T_{\mathrm{hidden}}$ that no agent sees, which in our evaluation is the hidden ProgramBench suite (Section~\ref{sec:eval-aqueduct}).
The resulting port becomes the reference for optimization.

\begin{figure}[t]
\centering
\includegraphics[width=\linewidth]{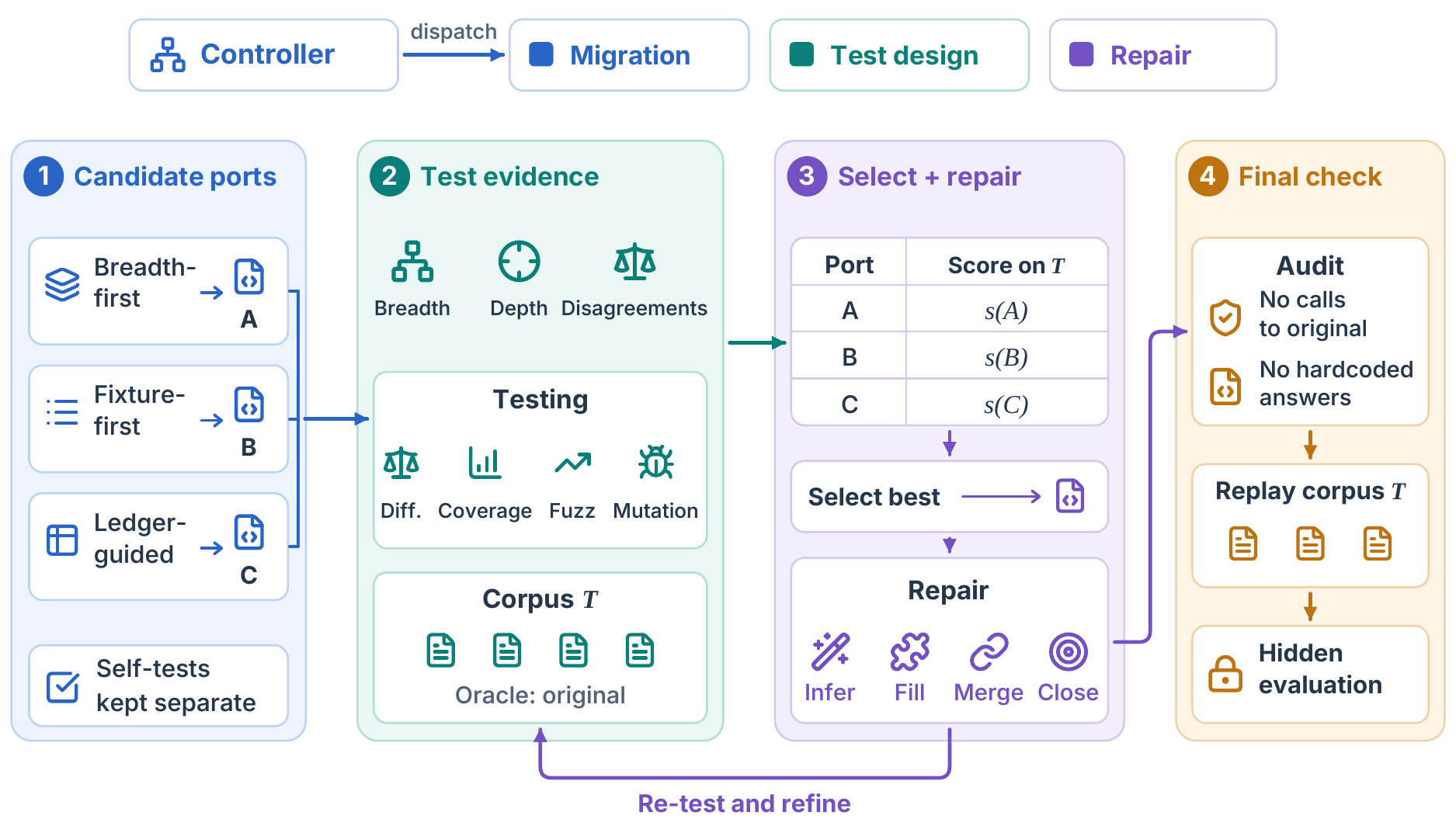}
\caption{\textbf{The \aqueduct{} migration workflow.} Three porting strategies produce candidate implementations. Test-design agents inspect the frozen candidates and build a shared corpus $T$, using differential execution, coverage, fuzzing, and mutation testing. The controller selects a port by agreement with the original across behavioral families. Repair agents revise the selected port, and rerunning the tests checks each revision as the corpus grows. The controller audits the resulting port for calls to the original and hardcoded answers before a single hidden evaluation.}
\label{fig:migration}
\end{figure}

\paragraph{Migration workflow.}
Agents do the work that requires judgment, and a deterministic controller makes every decision on the evidence (Figure~\ref{fig:migration}).
Three migration agents port each package under different strategies, so that their mistakes differ.
One ports the whole package at once, one first groups behavior into families such as a subcommand or an output format and builds test inputs for each family before implementing it, and one keeps an explicit ledger of every family and its status.
Once these candidates are frozen, three test-design agents extend the corpus toward uncovered entry points and error paths, structured and boundary inputs, and inputs on which the candidates disagree.
They rely on native Lean tools that we build.
A differential harness runs both programs on the same input and shrinks each divergence to a minimal counterexample, coverage instrumentation and coverage-guided fuzzing reach untested code, and faults that survive mutation testing mark behavior the corpus executes but does not check.
The controller replays the corpus on every candidate and selects the one that agrees most with the original, weighing agreement across behavioral families so that many easy cases cannot hide a missing command or format.
Repair agents then fix the selected port against its failing tests, and the controller keeps a fix only if it repairs a failure without breaking a passing test.
Part of the evidence is withheld from the repair agents, so every fix is checked on tests its author never saw.
Repair alternates with test generation, so each round faces a stricter corpus, until the port agrees with the original on the whole corpus or the budget runs out.
The controller finally audits the port for calls to the original and hardcoded answers and measures it once on $T_{\mathrm{hidden}}$.

\section{Evaluation}
\label{sec:eval}
\label{sec:eval-setup}

The study asks whether agents can find faster implementations under an equality-proof requirement and how the improvements affect applications.
Section~\ref{sec:eval-verina} measures how often agents succeed on the fixed set of 189 \verina{} functions and how their gains are distributed.
Section~\ref{sec:eval-aqueduct} examines whether local rewrites improve application runtime in the migrated repositories and how well the Lean ports preserve the original programs' behavior.
In both studies we use Codex with GPT-5.6 Sol at the \texttt{xhigh} reasoning setting, withhold the evaluation workloads from the optimizer, and require full equality to the reference, so the speedups measure improvements that generalize across inputs rather than specialization to the timing inputs.

\subsection{Function-Level Optimization on \verina{}}
\label{sec:eval-verina}

\paragraph{Setup.}
The study covers the 189 \verina{} references, comprising 108 basic and 81 advanced tasks.
Each function receives an isolated optimizer invocation containing the reference implementation and the equality theorem to prove.
The 54 functions whose initial candidates did not improve performance receive one fresh retry under the same conditions.
The timing corpus is frozen before optimization and contains 49,620 generated and official cases.
Timing variation can produce apparent improvements even when the program is unchanged.
PIE~\citep{shypula2024learning} reports a mean apparent speedup of 1.12$\times$, reaching 1.91$\times$ in the top 5\%, when timing 500 pairs of identical binaries.
We therefore use repeated paired trials.
The timing harness uses Lean~4.18.0 on an AMD EPYC 9654 processor and is pinned to one logical CPU.
After warm-up and iteration calibration, it runs nine paired trials, alternates execution order, and requires matching outputs.
Speedup is the baseline median time per corpus pass divided by the candidate median time per corpus pass.

For aggregate performance, we use the fastest verified implementation available for each function, including its reference.
Thus, each target contributes $\max(1,s_f)$ to the geometric mean and median, where $s_f$ is the best verified candidate's speedup defined in Section~\ref{sec:measurement}.

\begin{figure}[t]
\centering
\includegraphics[width=\linewidth]{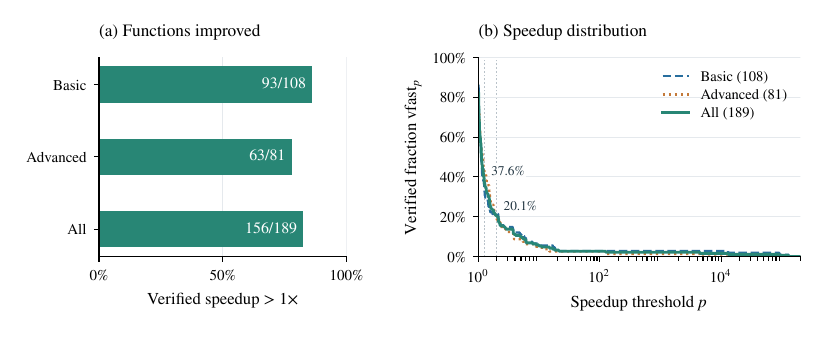}
\caption{\textbf{Function-level optimization on the 189 \verina{} references.} (a) The fraction of functions with a faster verified implementation, with counts shown inside the bars. (b) $\mathrm{vfast}_p$, the fraction with a verified speedup greater than threshold $p$ (Equation~\ref{eq:vfast}), using a logarithmic horizontal axis. Each curve covers every function in its suite.}
\label{fig:function-results}
\end{figure}

\begin{table}[t]
\centering
\caption{Verified optimization on the 189 \verina{} functions by suite. Aggregate speedups use the fastest verified implementation for each target, including its reference.}
\label{tab:verina-results}
\small
\begin{tabular}{lrrrr}
\toprule
Suite & Targets & Improved & Geom. mean & Median \\
\midrule
Basic & 108 & 93 & $1.959\times$ & $1.109\times$ \\
Advanced & 81 & 63 & $1.716\times$ & $1.163\times$ \\
All & 189 & 156 & $1.851\times$ & $1.117\times$ \\
\bottomrule
\end{tabular}
\end{table}

\paragraph{Agents improve 156 of the 189 functions.}
Agents produce faster implementations for 156 of the 189 targets (82.5\%), including 93 basic and 63 advanced functions (Figure~\ref{fig:function-results} and Table~\ref{tab:verina-results}).
The fresh retries contribute 21 of these improvements among the 54 retried functions.
The $\mathrm{vfast}_p$ curve in Figure~\ref{fig:function-results} shows how these gains are distributed.
At higher thresholds, 71 functions exceed $1.25\times$ speedup and 38 exceed $2\times$, giving $\mathrm{vfast}_{1.25}=37.6\%$ and $\mathrm{vfast}_2=20.1\%$.
Over the 189 targets, the resulting implementations achieve a geometric mean speedup of $1.851\times$ and a median of $1.117\times$.

\paragraph{Case studies show three ways an agent reaches a faster, provable implementation.}
The three largest speedups, up to $128{,}292\times$, replace iterated arithmetic sums with closed-form polynomials, and the rewrite of Section~\ref{sec:design} replaces recursion with dynamic programming.
Other cases show how the agent finds a rewrite it can prove.
First, the agent reuses verified library algorithms.
The reference for \code{SquareRoot} increments a candidate root until its square exceeds the input, and the rewrite calls Mathlib's logarithmic-time \code{Nat.sqrt}, running $21.16\times$ faster.
The proof shows by induction on the remaining bound that the loop returns \code{Nat.sqrt N}, maintaining that the candidate is at most the root and the root is at most the candidate plus the bound, and \code{isPerfectSquare} is rewritten the same way ($17.97\times$).
Second, the agent exploits properties of the problem.
The reference for \code{majorityElement} (\code{verina\_advanced\_35}) counts every element in a hash map and then searches the table for a count above half the length, while the rewrite returns as soon as one count crosses that threshold, running $3.74\times$ faster.
The early return is correct because at most one element can occur more than half the time, and the proof establishes this with a lemma that the counts of two distinct elements sum to at most the list length.
Third, the agent changes the representation while keeping the control flow.
The reference for \code{nthUglyNumber} builds its sequence by appending to a list and indexing into it, which takes quadratic time, and the rewrite performs the same steps on an array, running $124.90\times$ faster.
Because the two loops take identical steps, the proof only relates each array to the list it represents.

\subsection{Repository-Level Optimization on \aqueduct{} Ports}
\label{sec:eval-aqueduct}
\label{sec:eval-results}

\paragraph{Setup.}
To evaluate \aqueduct{}, we apply it to 40 repositories sampled from ProgramBench~\citep{yang2026programbench}, which supplies pinned C, C++, Go, and Rust programs with hidden end-to-end tests through their command-line interfaces.
The repositories range from 258 to 1,524,580 source lines and include archivers, compression libraries, build systems, code analyzers, and geospatial tools.
Agents may read the pinned source and execute the original, but cannot access the hidden tests or signals derived from them.
Of the 39,857 declared tests, the pinned originals pass 18,642 in the evaluation environment, and we score preservation of their behavior on this original-supported subset.
As a baseline, a single migration agent receives the same task and model configuration and produces a port directly.
The optimization study covers 17 of these ports, spanning C, C++, Go, and Rust programs from command-line utilities to a build system and a documentation generator, with 51 application scenarios, three per repository.
A separate agent constructs their function-level and application-level workloads from the public entry points and the migration test corpus, and the workloads are fixed before optimization.
For each repository, the optimizer inspects the source and call structure to select total functions without \texttt{IO} effects, then proposes replacements and proves them equivalent to the Lean references.
We compare migrated Lean, optimized Lean, and the original program on the independently constructed microbenchmarks and macrobenchmarks.
Each microbenchmark runs the three implementations on matched function inputs, and both kinds of benchmark follow the timing protocol of Section~\ref{sec:eval-verina}.
The macrobenchmarks exercise the application after integrating its accepted rewrites, so their speedups measure the combined effect on each scenario.

\begin{table}[t]
\centering
\caption{Performance of migrated and optimized applications at both scales. Speedup is baseline runtime divided by evaluated runtime, and values above one favor the evaluated implementation. Faster counts the kernel-checked rewrites or application scenarios whose speedup exceeds one.}
\label{tab:software-results}
\small
\setlength{\tabcolsep}{4pt}
\begin{tabular}{lrrrrrr}
\toprule
& \multicolumn{3}{c}{Function level (92 rewrites)} & \multicolumn{3}{c}{Application level (51 scenarios)} \\
\cmidrule(lr){2-4}\cmidrule(lr){5-7}
Comparison & Faster & Geom. mean & Median & Faster & Geom. mean & Median \\
\midrule
Optimized Lean vs. migrated Lean & 92/92 & $1.864\times$ & $1.402\times$ & 47/51 & $1.120\times$ & $1.077\times$ \\
Optimized Lean vs. original & 15/92 & $0.235\times$ & $0.217\times$ & 3/51 & $0.148\times$ & $0.160\times$ \\
Migrated Lean vs. original & 12/92 & $0.126\times$ & $0.120\times$ & 3/51 & $0.132\times$ & $0.158\times$ \\
\bottomrule
\end{tabular}
\end{table}

\begin{figure}[t]
\centering
\includegraphics[width=\linewidth]{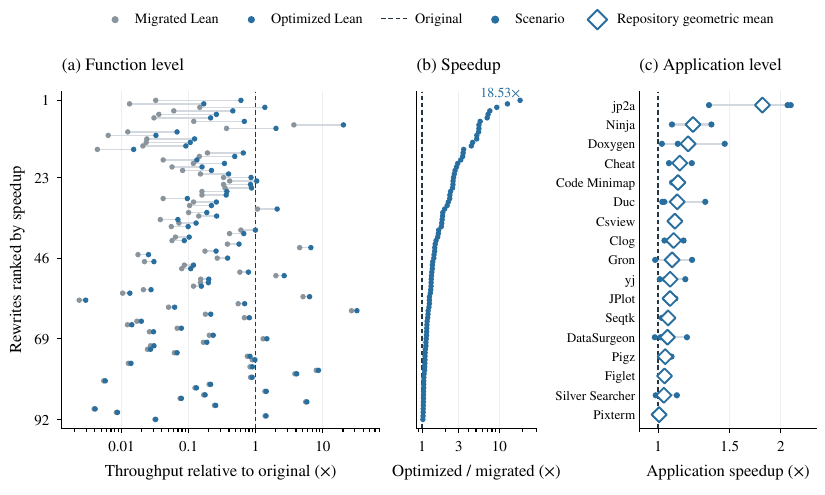}
\caption{\textbf{Repository-level optimization measured at two scales.} (a) Each row is one of the 92 accepted rewrites, ranked by its speedup over the migrated Lean reference, with the migrated and optimized implementations placed by throughput relative to the original source-language operation. The dashed line marks the original. (b) The speedup of each rewrite over its migrated reference, on the same rows. (c) Application speedup of optimized over migrated Lean for the 51 scenarios, grouped by repository, with diamonds marking each repository's three-scenario geometric mean.}
\label{fig:repository-results}
\end{figure}

\paragraph{Agents produce substantial speedups with proved equivalence.}
The agents produce 92 proved rewrites, and each is faster than the migrated declaration it replaces (Figure~\ref{fig:repository-results}a and Table~\ref{tab:software-results}).
The geometric mean speedup is $1.864\times$, the median is $1.402\times$, and the maximum is $18.53\times$.
The rewrites fall into three main patterns.
\begin{enumerate}[leftmargin=*]
    \item \emph{Algorithmic improvement.} The rewrite replaces the algorithm with one of lower complexity, mainly by removing loops. One migrated function deduplicates a list by filtering and rebuilding it after every insertion, which is quadratic, and the rewrite produces the same list in a single linear pass with a hash set. Another normalizes the same query strings once per item inside a loop, and the rewrite normalizes them once before the loop.
    \item \emph{Representation improvement.} The rewrite keeps the algorithm but computes over a cheaper data representation. One rewrite validates a string in place instead of first converting it to a character list, another folds over an array directly instead of converting it to a list, and a third reserves the final buffer size before appending bytes instead of growing the buffer element by element.
    \item \emph{Evaluation-order improvement.} The rewrite keeps both the algorithm and the representation but reorders the computation so that inputs the source code suggests are common take the cheap path. One rewrite tests the frequent ASCII cases before falling back to general Unicode predicates, another selects the comparison function once before sorting instead of re-branching on every comparison, and a third returns immediately on a zero exponent instead of entering the multiplication loop.
\end{enumerate}
In 15 of the 92 functions, the optimized Lean declaration is faster than the matched original operation, compared with 12 for the migrated declarations.
The port overtakes the original when the original performs the same avoidable work, as when the per-item normalization above is hoisted out of the loop and the Lean function goes from $0.37\times$ to $2.01\times$ the speed of the Go original.
In the remaining functions, the rewrites narrow the gap to the original but do not close it.
Appendix~\ref{app:casestudies} examines five representative cases.

\paragraph{The function-level gains survive integration but shrink.}
The optimized executables are $1.120\times$ faster than the migrated executables, faster in 47 of 51 scenarios, and faster in all 17 repositories by the per-repository geometric mean, yet the improvement is far below the $1.864\times$ function-level gain (Figure~\ref{fig:repository-results}c).
The application speedup is bounded by the share of scenario time spent in the rewritten functions, and the rest of each scenario, including its I/O, runs unchanged.
Because every integrated rewrite is proved equal to the declaration it replaces, the optimized applications compute the same results as the migrated ones.
Closing the remaining end-to-end gap requires optimizing beyond isolated functions, through module-level rewrites of shared representations (Section~\ref{sec:formulation}) and, for the I/O paths, the extension to opaque effects (Section~\ref{sec:future}).

\paragraph{\aqueduct{} recovers more source-program behavior than a single agent.}
To assess the Lean references supplied to repository-level optimization, we compare their behavior with the original programs on the hidden ProgramBench tests.
Across the 40 repositories, the \aqueduct{} ports pass 16,345 of the 18,642 original-supported tests, compared with 12,709 for the single-agent baseline.
These counts correspond to recovery rates of 87.68\% and 68.17\%, a difference of 19.50 percentage points.
The per-repository comparison shows where the difference comes from.
On nine repositories, the single agent's port passes none of the tests, while the \aqueduct{} port recovers between 61.10\% and 100\%, and these repositories account for 2,685 of the net gain of 3,636 tests.
On the twelve other repositories where \aqueduct{} improves, the single agent already reproduces the common core of the program, and the additional 960 tests come from the long tail of rare option combinations, malformed and boundary inputs, exact diagnostics and exit conventions, and environment-dependent behavior that no single porting attempt covers completely.
On the remaining repositories the two methods mostly tie at high recovery, and the single agent leads on four by at most five tests.
\aqueduct{} addresses both the ports that fail outright and the long tail.
Independent candidates fail on different behaviors, their disagreements become tests, and the tests drive selection and repair.
The additional tests generated during migration also detect more source-level mutants than the ProgramBench suite in a paired comparison (Appendix~\ref{app:migration}).
Removing candidate diversity, test-generation feedback, or repeated repair reduces recovery in the component study (Appendix~\ref{app:ablation}).

\section{Research Directions}
\label{sec:future}

The verified improvements in Section~\ref{sec:eval} range from removing repeated work and adding fast paths to changing algorithms and representations.
They leave open how agents should organize this search, which improvements matter in an application, and how the task extends to more software.

\paragraph{Allocating effort between implementation and proof.}
A faster candidate may require an invariant that is difficult to discover, whereas a smaller change may admit a short proof.
Future studies can compare optimizing before proving, proving after each change, and interleaving implementation and proof construction under matched time and token budgets.
Tracking verified speedup throughout a run would measure how quickly each strategy finds a useful replacement and whether further search improves it.
Matched runs with and without a proof requirement would show how the requirement changes which transformations agents attempt.
Comparing agent models and search strategies under matched budgets would likewise show how verified speedup depends on proving capability.

\paragraph{Reusing proofs across optimizations.}
The Gron example reuses earlier lemmas to justify a change in string representation.
This raises the question of when an agent should invest in a general lemma rather than prove only the current rewrite.
Comparing fresh proof search with access to lemmas from earlier optimizations would measure the effect on proof cost and verified improvements.
Separating reuse within a repository from transfer across repositories would show how much of the resulting proof knowledge applies beyond the original program.
Reuse during software maintenance introduces a further question, since callers and dependencies can change after a rewrite is proved.
Studying this setting would test whether agents can adapt earlier optimizations and proofs as an application evolves.

\paragraph{Selecting targets for application performance.}
The difference between function-level and application-level gains makes target selection a central part of repository-level optimization.
A direct next experiment is to compare selection from source inspection with selection informed by profiles on development workloads.
A second is to compare local function rewrites with module-level rewrites of a shared representation under matched budgets, measuring proof effort against application-level gain.
Evaluation on separate application workloads can then test whether those choices improve total runtime.
Comparing individual rewrites with their combinations would expose interactions that isolated microbenchmarks do not measure.

\paragraph{Specializing to a kind of workload.}
Our study withholds the evaluation workload and requires full equality, so every accepted rewrite is correct on all inputs and none is tuned to the timing inputs.
Describing the kind of workload a function serves, or providing a sample of it, would let agents specialize data representations and control paths to that workload, as Jitskit~\citep{liu2026time} does for whole systems.
Section~\ref{sec:formulation} shows that the task supports this without giving up a machine-checked correctness argument, through a guarded replacement or the relaxed obligation $\forall x,\, P(x) \rightarrow f(x)=f'(x)$.
Measuring how much additional speedup specialization yields, and how the choice of the predicate $P$ trades proof effort against coverage of the workload, is a direct extension of the present study.

\paragraph{Learning from optimization trajectories.}
Optimization trajectories connect candidate programs and measured performance with proof states and verifier feedback.
Training on these trajectories could teach an agent which changes to attempt, when to continue a proof, and when to revise an implementation.
Unsuccessful attempts could help it distinguish transformations that need a different invariant from those that need a different implementation.
Evaluation on held-out repositories under fixed budgets would test whether this training improves verified speedup beyond the programs represented in the training data.

\paragraph{Extending equivalence beyond equality.}
Our study covers functions whose effects are expressed inside Lean, that is, pure functions and monadic code over Lean-defined monads such as state and exceptions, because equality of the definitions is then the complete correctness relation and Lean's kernel checks it directly.
Equality cannot be proved directly for opaque effects, such as file and network access, whose behavior Lean does not define.
They fit the same task once they are modeled.
Such operations can be abstracted behind an interface of primitive operations, so that a program becomes a pure function from the responses it receives to the requests it issues and the results it returns, and equivalence becomes equality of those functions.
Concurrency requires a relation over the interleavings or traces that callers can observe.
Equality can also be stronger than callers need.
A floating-point result that callers use within a tolerance admits rewrites that change rounding, such as reassociating a sum.
The obligation then becomes $\forall x,\ R(f(x), f'(x))$ for a relation $R$ such as a bound on relative error, as with the refinement relation that Section~\ref{sec:formulation} requires when a new representation appears in exported types.
Each relation fixes which observations a rewrite must preserve and which it may change, such as the order of independent operations, the amount of buffering, or the rounding of a result, and choosing it is part of stating the task for a given codebase.
A relation weaker than equality is a partial specification and carries the risk that Section~\ref{sec:formulation} describes, since a loose relation lets a rewrite drop behavior that callers rely on, so it should follow from what callers observe.
Optimizing the I/O paths of the applications in Section~\ref{sec:eval} requires this extension.

\paragraph{Extending the optimization space.}
The rewrites in Section~\ref{sec:eval} change algorithms and data representations at the Lean source level and run on Lean's reference-counted runtime~\citep{ullrich2019counting}, and they remain slower than the native programs they were migrated from (Table~\ref{tab:software-results}).
The equality obligation does not depend on the level at which the replacement is written, so a rewrite could target explicit memory layouts, unboxed representations, or vector instructions, with a Lean-checked proof that the compiled artifact implements the reference function and the trusted boundary moved to that program's lowering.
Much of the Lean compiler and its standard library are Lean programs as well, so verified optimizations of core library functions, registered as \texttt{@[csimp]} replacements, would carry their proofs to every caller, and optimizations of compiler passes would extend agent-built verified compilation~\citep{rinard2026testing} to performance work, so that agents would improve the system on which their own rewrites run.

\paragraph{Certifying migration.}
For migrated applications, the original program supplies a further correctness relation to establish.
A proof connecting that program to its Lean port could compose with the optimization proof to certify the full path from the original to the optimized implementation.
Constructing this proof requires models of the source semantics, memory, and runtime operations that connect the two languages.
A shorter route exists where such models are available.
Aeneas~\citep{ho2022aeneas} translates a subset of Rust into pure Lean functions, so optimizing a Rust function and proving the Lean models of the two versions equal would apply the task to code that runs natively, with the translation as the trusted step.

\paragraph{Proving performance properties.}
The objective in this paper measures execution time on a fixed workload, and the proof says nothing about speed.
Asymptotic bounds can be proved in Lean, and CSLib~\citep{barrett2026cslib} treats time complexity as a computational effect, so that a computation returns its value together with a cost under a declared cost model and correctness and complexity are proved separately.
A proved improvement in a bound can still run slower in practice when its constants are large.
An explicit cost model~\citep{shyamsunder2026formal} would let a rewrite carry both, a proof that it reduces a resource measure or improves an asymptotic bound and its measured speedup, with the agent relating the original and rewritten computations through their outputs and their costs.
Comparing proved bounds with measured runtime across rewrites would show which cost models guide practical optimization and where constants dominate.

\section{Conclusion}
\label{sec:conclusion}

We formulated agentic verified optimization as a search for a faster implementation and a proof of equivalence to an existing program.
Fixing the reference gives the agent a precise correctness obligation while leaving it free to remove work, add fast paths, or replace the algorithm.
The study finds verified improvements in standalone Lean functions and in applications migrated by \aqueduct{}.
Agents are capable of producing verified optimizations, including rewrites that eliminate intermediate structures or reorder computation.
Application-level gains are smaller than function-level gains, and four scenarios do not improve, so a locally faster function does not guarantee a faster application.
Target selection should therefore account for each function's share of application runtime.
Together, the formulation and findings support studying how agents coordinate implementation and proof search to improve application performance with machine-checked correctness guarantee.

\clearpage
\appendix
\section{Correctness Checks}
\label{app:acceptance}

The protocol applies to functions whose effects are expressed inside Lean, including pure functions and computations in monads such as \texttt{Id}, \texttt{StateM}, and \texttt{Except}, and excludes functions whose behavior depends on opaque effects such as \texttt{IO} or foreign calls.
For these functions, the equality theorem covers every input without a bound on input size or recursion depth.
Following Vero~\citep{ye2026vero}, the acceptance checker extracts the submitted declarations and inserts them into a fresh copy of the reference environment.
It then applies four checks to protect the reference, validate the proof, and ensure that the executable replacement is the definition being proved.
\begin{enumerate}[leftmargin=*]
\item \textbf{Build.} In the fresh environment, every existing declaration of $\mathcal{R}$, including the reference $f$ and its callers, remains byte-identical, and the submission adds only the replacement $f'$, its auxiliary definitions and lemmas, and the proof $\pi$. In a separate integration check, the acceptance checker registers the equality theorem as a \texttt{@[csimp]} replacement, so compiled callers use $f'$, and checks that the resulting repository $\mathcal{R}'$ builds with all submitted declarations.
\item \textbf{Axioms.} The axioms used by $\pi$ and every auxiliary declaration, including through their dependencies, must be a subset of $\{\texttt{propext}, \texttt{Quot.sound}, \texttt{Classical.choice}\}$. In particular, \texttt{sorryAx} is forbidden.
\item \textbf{Executability.} The rewrite and its agent-supplied computational dependencies must be computable, without \texttt{partial} or \texttt{unsafe} definitions or \texttt{@[implemented\_by]} and \texttt{@[extern]} substitutions. They use the fixed library and runtime supplied with the task.
\item \textbf{Kernel-only checking.} The proof must not use \texttt{native\_decide}, which appeals to compiled execution to discharge a proof step.
\end{enumerate}
Condition 3 prevents executable substitutions for the definition being proved, while Condition 4 requires proof steps to be checked by the kernel.
The protocol also rejects agent-introduced typeclass instances that would change the interpretation of equality for the types involved.
The checked proof and its dependencies can be replayed in Lean.
Accepted implementations are compiled and timed with the task's fixed Lean compiler, standard library, and runtime.

\section{Migration Pipeline and Results}
\label{app:migration}

\paragraph{Pipeline.}
The migration workflow produces the Lean references used by the optimizer.
Its three candidate strategies emphasize broad interface coverage, executable fixtures, and an explicit ledger of behavioral obligations.
Separate test-design agents inspect frozen candidates and generate inputs for uncovered behavior, boundary cases, and disagreements between ports.
The controller compares each port with the original on those inputs, selects a candidate, and coordinates repair.
Coverage and mutation feedback guide further test generation.
Migration runs in the benchmark's white-box setting.
Agents may read the pinned source and execute the original program, but the hidden tests and any signal derived from them stay inaccessible throughout, so the scores measure generalization to behavior that no agent observed.
We compare \aqueduct{} against a baseline of one migration agent given the source repository and the migration task, which represents what one-shot agentic migration achieves without the pipeline.
Both run in the same setting with the same model.

\paragraph{Interface comparison.}
The same comparison rule applies to the generated corpus $T$ during migration and to $T_{\mathrm{hidden}}$ at the final evaluation.
Unstable behavior, including randomness, host-dependent floating-point formatting, and language-specific error text, is excluded from that interface or handled by its declared comparison rule.
Applying the same comparison rule throughout candidate selection, repair, and hidden evaluation keeps the required behavior consistent across the migration process.

\paragraph{Recovery.}
ProgramBench declares 39,857 tests across the 40 sampled repositories, of which the original programs pass 18,642 in the evaluation environment.
Recovery is the fraction of these original-supported tests that the port passes.
Across these repositories, the \aqueduct{} ports pass 16,345 tests and the single-agent baseline passes 12,709, or 87.68\% and 68.17\% recovery.

\paragraph{Generated-test quality.}
The pipeline rests on the premise that generated evidence must discriminate, telling faithful ports apart from unfaithful ones and exposing the faults a weaker suite would miss.
We evaluate this premise directly with mutation analysis.
After the test suites are frozen, we inject deterministic source-level faults into the original programs, rebuild each mutated checkout, and count only mutants that compile.
A mutant is killed only when a test that passes on the unmutated program fails on the mutant.
Because the faults are planted after freezing, the audit is independent of the migration search that produced the tests.
The test-design agents generated 33,671 behavioral tests across the 40 migration targets.

\emph{Generated tests versus the benchmark's own tests.}
The central question is whether the generated suite observes behavioral differences that the external benchmark does not, so we score the same compiled mutants under both the generated tests and the hidden ProgramBench tests.
On 36 compiled mutants across 9 repositories, the generated tests kill 28 (77.8\%) against 17 (47.2\%) for ProgramBench (Table~\ref{tab:mutants}).
The asymmetry is sharpest among mutants that exactly one suite catches.
The generated tests account for 12 of these 13, ProgramBench for 1 (exact paired binomial $p = 0.0034$).
The generated suite discriminates behavioral differences that the benchmark's own tests do not observe, which is why the pipeline generates its evidence rather than borrowing it.

\begin{table}[h]
\centering
\caption{Mutants killed by the generated tests and by the hidden ProgramBench tests on the same 36 compiled mutants across 9 repositories.}
\label{tab:mutants}
\small
\begin{tabular}{lr}
\toprule
Outcome & Mutants \\
\midrule
Killed by both suites & 16 \\
Killed by generated tests only & 12 \\
Killed by ProgramBench tests only & 1 \\
Killed by neither & 7 \\
\midrule
Total scored & 36 \\
\bottomrule
\end{tabular}
\end{table}

\emph{Auditing the suite against itself.}
We also measure the generated suite's absolute kill rate on the full audit population (Table~\ref{tab:mutation-audit}).
The 95\% interval is a repository-cluster bootstrap.
These are end-to-end behavioral tests at the program boundary, where a source-level mutant registers only if its effect propagates to observable output, so interface-level kill rates run below what unit-level mutation testing reports.
Within the pipeline, each surviving mutant marks a behavior that the corpus executes but does not yet constrain, and it becomes a target for the next round of test generation.
The behaviors the suite does not yet check are therefore enumerated, as in the 7zip case study, which itemizes the port's remaining obligations.

\begin{table}[h]
\centering
\caption{Mutation audit of the generated suite on the full audit population.}
\label{tab:mutation-audit}
\small
\begin{tabular}{rrrrrr}
\toprule
Repositories & Compiled mutants & Killed & Score & Median per repository & 95\% interval \\
\midrule
25 & 190 & 101 & 53.2\% & 50.0\% & 39.1\% to 67.0\% \\
\bottomrule
\end{tabular}
\end{table}

\section{Migration Ablations}
\label{app:ablation}

\begin{figure}[t]
\centering
\includegraphics[width=\linewidth]{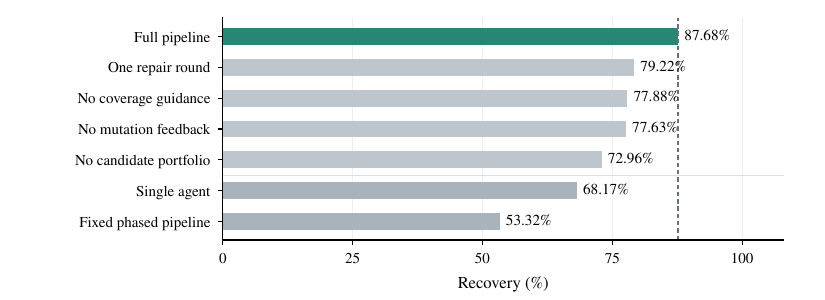}
\caption{\textbf{Contribution of migration components.} Recovery under component ablations and two pipeline baselines. Removing candidate diversity produces the largest decrease among the four component ablations.}
\label{fig:migration-ablations}
\end{figure}

Figure~\ref{fig:migration-ablations} compares the full pipeline with four component ablations and two baselines.
\emph{One repair round.} The pipeline runs candidate generation, test design, and selection in full, but performs a single round of repair instead of alternating repair with evidence generation until convergence. This measures the full repair loop.
\emph{No coverage guidance in the testing loop.} The testing loop loses coverage instrumentation, so test generation no longer learns which code paths remain unexercised. This measures the coverage signal.
\emph{No mutation feedback in the testing loop.} The testing loop loses mutation testing, so faults the corpus fails to catch no longer become targets for the next round of test generation. This measures the mutation signal.
\emph{No candidate portfolio (best single strategy).} A single migration agent ports each package under one fixed strategy instead of three. We run each of the three strategies this way and the figure reports the best one. This measures candidate diversity and selection together.
\emph{Baseline (single agent, no pipeline).} A plain agent ports each package.
\emph{Fixed phased pipeline.} A variant with the same migration and testing tools exposes them as a fixed sequence of phases. This measures the value of letting agents revisit earlier steps rather than follow a fixed order.

\paragraph{Each component contributes to recovery.}
The full pipeline recovers 87.68\% of the original behavior against 68.17\% for a single agent with the same source and toolchain, and every component ablation falls between the two, while the fixed phased pipeline falls below the single agent.
The following paragraphs attribute this gap to individual components.

\paragraph{Iterative repair outperforms a fixed sequence of phases.}
The phased pipeline recovers less than even the single-agent baseline, and it contains the same tools as the full pipeline, so the tools are not the bottleneck.
Its phases run in a fixed order and never revisit earlier decisions, so when a later phase uncovers behavior the port misses, the agent cannot return to update the behavioral surface it committed to earlier.
The agents fill gaps in their ports when they can revisit earlier decisions, but not when a fixed sequence prevents it.
This result suggests giving agents freedom over how they work and concentrating rigor in the acceptance criteria and the checks that judge the result.

\paragraph{Candidate diversity provides the largest gain.}
Committing to the best single migration strategy caps recovery at 72.96\%.
The controller picks the first strategy for 18 repositories, the second for 20, and the third for 2, so no strategy dominates and the gain comes from having a portfolio and the evidence to choose among it.
This also reflects a structural property of the search.
Repair is incremental and rarely changes a port's overall structure, so the structural diversity that matters must already exist among the initial candidates.

\paragraph{Both testing signals improve the corpus in complementary ways.}
Removing coverage guidance or mutation feedback costs 9.80 and 10.05 points, more than the 8.46 points lost by restricting repair to a single round, so the quality of the test corpus matters more than the repair loop that consumes it.
The two signals match their designed roles.
Coverage guidance steers test generation toward behavior no test has exercised, and mutation feedback hardens the checks on behavior the corpus already reaches.
Repair can only fix what a test exposes, so a weaker corpus leaves divergences undetected and the pipeline treats them as agreement.

\section{Additional Case Studies}
\label{app:casestudies}

\subsection{Longest Increasing Subsequence}
\label{app:lis-proof}

The following theorem and proof establish equivalence for the rewrite in Figure~\ref{fig:task-format}.
The input \code{nums} is an arbitrary list of integers, so the theorem covers every such list.
The proof quantifies over \code{prev}, the optional preceding element used by the recursive helper, and shows that querying the computed table returns the same length as that helper.

\noindent\begin{minipage}{\linewidth}
\begin{lstlisting}[language=lean,firstline=72,lastline=102,firstnumber=1]
theorem longestIncreasingSubsequence_optimized_eq_baseline (nums : List Int) (h_precond : longestIncreasingSubsequence_precond nums) :
    longestIncreasingSubsequence_optimized nums h_precond = longestIncreasingSubsequence nums h_precond := by
  unfold longestIncreasingSubsequence_optimized longestIncreasingSubsequence
  clear h_precond
  suffices h : ∀ prev, lisTableQuery (lisTable nums) prev =
      longestIncreasingSubsequence.helper (fun a b => if a > b then a else b) nums prev by
    exact h none
  have max_eq (a b : Nat) : Nat.max a b = (if a > b then a else b) := by
    rw [Nat.max_eq_max]
    by_cases hab : a > b
    · rw [if_pos hab, Nat.max_eq_left (Nat.le_of_lt hab)]
    · rw [if_neg hab, Nat.max_eq_right (Nat.le_of_not_gt hab)]
  induction nums with
  | nil => intro prev; rfl
  | cons x xs ih =>
      intro prev
      cases prev with
      | none =>
          simp only [lisTable, lisTableQuery, longestIncreasingSubsequence.helper]
          rw [ih (some x), ih none]
          exact max_eq _ _
      | some p =>
          simp only [lisTable, lisTableQuery, longestIncreasingSubsequence.helper, Option.some.injEq, reduceCtorEq, ↓reduceIte,
            Option.get!, Bool.if_true_right]
          by_cases hp : p < x
          · simp only [hp, ↓reduceIte, Bool.true_eq]
            rw [ih (some x), ih (some p)]
            exact max_eq _ _
          · simp only [hp, ↓reduceIte, Bool.false_eq]
            rw [ih (some p)]
            simp
\end{lstlisting}
\end{minipage}

Across the 261 inputs in the evaluation corpus, median time per corpus pass falls from $1.424$ seconds to $318.653$ microseconds, a speedup of $4{,}467.871\times$.

\subsection{Migration at Scale: 7zip}

7zip is the fourth largest repository in our sample at 252,968 lines of source, and it tests whether the pipeline can expand a port across a large behavioral surface for frequently used real-world software.
ProgramBench declares 883 original-supported tests for 7zip, spanning more than a dozen archive formats and their command-line, stream, and filesystem interactions.
The final port recovers 554 of them (62.74\%), against 28.77\% for the baseline, a direct measure of where migration is hardest.
The port is a substantial native Lean implementation.
It handles ZIP, TAR, 7z, bzip2, Zstandard, raw LZMA and LZMA86, split archives, AR and DEB, CPIO, RPM, ISO-9660 with Joliet, and CAB with None and MSZip compression, and it implements listing, testing, extraction, creation, update and delete paths, include and exclude filters, standard-stream routing, overwrite behavior, and hash manifests.
No single porting attempt produces this surface.
It accumulates through the search, as test-design agents surface unhandled formats and options, each divergence becomes a repair target, and each repaired behavior freezes into a permanent regression case.
At the final checkpoint the port passes 541 of 579 generated behavioral tests, and the search's inventory of behavioral obligations still lists 938 open items when the budget is exhausted, behaviors the corpus has named but not yet established evidence for.
These open obligations account for the gap between the generated-corpus pass rate and hidden-set recovery.
The port's remaining gaps are thus measured and itemized, and they direct the next round of the search.

\subsection{Optimization Case Studies}

We examine five proved optimizations that illustrate the three rewrite patterns of Section~\ref{sec:eval-aqueduct}.

\paragraph{Algorithmic improvement (Cheat \texttt{findConsolidated}).}
The migrated function first builds the complete consolidated sheet list and then searches it for the requested title, so every lookup pays for constructing entries it immediately discards.
The optimized declaration searches from the end of the ordered paths and returns the last matching title directly, never materializing the intermediate list.
The equality theorem proves that direct reverse search returns exactly the element the build-then-search pipeline would find, folding the consolidation semantics into the lookup.
The migrated declaration is already faster than the matched Go operation, and the rewrite, $5.46\times$ faster than migrated Lean, widens that margin to $20.36\times$ by removing the build-then-search work that the Go code also performs.

\paragraph{Algorithmic improvement (Cheat \texttt{consolidatedSheets}).}
The migrated function builds its result by filtering the entire accumulated list before every append, repeatedly traversing and reallocating a growing structure.
The optimized declaration walks the input once in reverse, tracks seen titles in a hash set, and constructs the output list once.
The equality theorem proves that the reverse traversal preserves the exact last-title-wins semantics of the original fold, a property that is easy to get wrong in a manual rewrite.
The rewrite is $1.46\times$ faster than migrated Lean and $6.65\times$ faster than the matched Go operation.

\paragraph{Representation improvement (Gron \texttt{validIdentifier}).}
The migrated function materializes a full character list to check an identifier.
The optimized declaration reads the first character directly and checks the remainder through a string slice, allocating nothing.
The theorem composes the slice-to-list equivalence with previously proved character-classification replacements, so later proofs reuse lemmas from earlier rewrites.
The rewrite is $1.96\times$ faster than migrated Lean and $2.10\times$ faster than Go.

\paragraph{Algorithmic improvement (Cheat \texttt{filterSheets}).}
The migrated function re-trims every requested tag once per sheet.
The optimized declaration normalizes the tags once and reuses the prepared list across all sheets, changing repeated normalization into a single pass.
The rewrite is $5.45\times$ faster than migrated Lean and $2.01\times$ faster than the matched Go operation, and the output checksums agree across all three implementations.

\paragraph{Evaluation-order improvement (Code Minimap \texttt{pow10}).}
The migrated helper enters its multiplication loop even for a zero exponent.
The optimized declaration returns the zero case immediately, producing a $9.29\times$ within-Lean speedup and running $1.37\times$ faster than the Rust original on a microbenchmark whose inputs include the zero exponent.
The equality theorem shows that the fast path changes no result, because the declarations with and without the fast path are equal on every input.
Under the equivalence obligation, the agent can therefore add fast paths for inputs it expects to be common without restricting correctness to those inputs.

In all five cases, the rewrite removes a pattern common in migrated code, such as repeated filtering, intermediate materialization, redundant normalization, or a missed fast path, and the equality proof guarantees that removing the pattern preserves behavior on every input.

\end{document}